\documentstyle[12pt]{article}
\begin{document}

\begin{titlepage}
\begin{center}
\today     \hfill    LBNL-40024 \\
~{} \hfill UCB-PTH-97/08  \\

\vskip .25in

{\large \bf Neutrino physics from a U(2) flavor symmetry}\footnote{This work
was supported in part by the Director, Office of Energy Research, Office of
High Energy and Nuclear Physics, Division of High Energy Physics of the U.S.
Department of Energy under Contract DE-AC03-76SF00098.  LJH was also supported
in part by the National Science Foundation under grant PHY-95-14797.}

\vskip 0.3in

Christopher D. Carone$^1$ and Lawrence J. Hall$^{1,2}$

\vskip 0.1in

{{}$^1$ \em Theoretical Physics Group\\
     Ernest Orlando Lawrence Berkeley National Laboratory\\
     University of California, Berkeley, California 94720}

\vskip 0.1in

{{}$^2$ \em Department of Physics\\
     University of California, Berkeley, California 94720}
        
\end{center}

\vskip .1in

\begin{abstract}
We consider the neutrino physics of models with a sequentially 
broken U(2) flavor symmetry.  Such theories yield the observed pattern of 
quark and lepton masses, while maintaining sufficient degeneracies between
superparticles of the first two generations to solve the supersymmetric 
flavor problem.  Neutrino mass ratios and mixing angles in these
models may differ significantly from those of the charged leptons,
even though the neutrinos and charged leptons transform identically under the 
flavor group.  A wide class of well-motivated U(2) theories yield order 
one $\nu_\mu$-$\nu_\tau$ mixing, without a fine-tuning of parameters.  These 
models provide a natural solution to the atmospheric neutrino deficit, 
and also have distinctive signatures at long-baseline neutrino 
oscillation experiments.
\end{abstract}

\end{titlepage}
\renewcommand{\thepage}{\roman{page}}
\setcounter{page}{2}
\mbox{ }

\vskip 1in

\begin{center}
{\bf Disclaimer}
\end{center}

\vskip .2in

\begin{scriptsize}
\begin{quotation}
This document was prepared as an account of work sponsored by the United
States Government. While this document is believed to contain correct 
information, neither the United States Government nor any agency
thereof, nor The Regents of the University of California, nor any of their
employees, makes any warranty, express or implied, or assumes any legal
liability or responsibility for the accuracy, completeness, or usefulness
of any information, apparatus, product, or process disclosed, or represents
that its use would not infringe privately owned rights.  Reference herein
to any specific commercial products process, or service by its trade name,
trademark, manufacturer, or otherwise, does not necessarily constitute or
imply its endorsement, recommendation, or favoring by the United States
Government or any agency thereof, or The Regents of the University of
California.  The views and opinions of authors expressed herein do not
necessarily state or reflect those of the United States Government or any
agency thereof, or The Regents of the University of California.
\end{quotation}
\end{scriptsize}

\vskip 2in

\begin{center}
\begin{small}
{\it Lawrence Berkeley Laboratory is an equal opportunity employer.}
\end{small}
\end{center}

\newpage
\renewcommand{\thepage}{\arabic{page}}
\setcounter{page}{1}
\section{Introduction} \label{sec:intro} \setcounter{equation}{0}

The question of neutrino masses is one of the most interesting in particle
physics -- especially in view of persistent observations suggesting non-zero
neutrino masses and mixing angles. It is straightforward to construct 
models for neutrino masses, and it is certainly very easy to understand 
why neutrino masses are much lighter than the charged leptons and quarks. 
However, the predictions for neutrino masses and mixing angles depend 
strongly on theoretical assumptions, since the neutrino mass matrix
has a symmetry structure which is very different from that of the charged 
fermions\footnote{The flavor symmetry group of the leptons of the standard 
model is $U(3)_L \times U(3)_R$, with the left(right)-handed leptons 
transforming as $[3,1]$ ($[1,3]$).  While the charged lepton masses 
transform as $[3,3]$, the neutrino masses transform as $[6,1]$.}. This 
leads to a decoupling of the unknown neutrino masses from the known 
charged fermion masses.  For example, one factor which determines the 
overall neutrino mass scale is the breaking of lepton number, about 
which we have no experimental information. We will not attempt to predict 
this overall scale of neutrino masses. In this paper we are able to make 
predictions for neutrino mass ratios and mixing angles by assuming a 
symmetry which re-couples, to a large degree, the neutrino and charged 
fermion masses.

Neutrino masses are part of the larger question of flavor physics -- how 
are the flavor symmetries of the standard model gauge interactions broken 
to yield fermion masses and mixing angles? In supersymmetric theories, 
flavor physics becomes much richer as the squarks and sleptons must also 
have mass matrices. Furthermore, flavor physics becomes constrained in 
new ways, because some form of ``super-GIM'' mechanism is necessary to 
suppress the flavor changing neutral current effects induced by 
supersymmetric gauge interactions. 

An approximate flavor $U(2)$ symmetry has recently been proposed as a 
simple and economical framework for understanding flavor in supersymmetric 
theories \cite{u212,u23}. The idea is that $U(2)$, and its breaking 
pattern, provide a basic order to the spectrum of quarks, leptons, and 
their superpartners, in the same way that the $SU(2)$ isospin symmetry 
provided for nuclear states, and the flavor $SU(3)$ provided for 
hadronic physics. The small values of the light quark and lepton masses 
are governed by two small $U(2)$ symmetry breaking parameters, as are the 
small CKM mixing angles. The same two symmetry 
breaking parameters are also responsible for the small non-degeneracies 
among the squarks and sleptons -- leading to a ``super-GIM'' mechanism. 
The choice of $U(2)$ transformations for the symmetry breaking parameters 
leads to relations between the CKM mixing angles and ratios of quark masses --
these are the analogue of the Gell-Mann-Okubo mass relation. These 
relations are in agreement with current measurements, and will be more 
precisely tested by future measurements \cite{BHR97}. The $U(2)$ theory 
also provides significant motivation for pursuing searches for $\mu 
\rightarrow e$ conversion and for electric dipole moments of the electron 
and neutron. The virtual effects of superpartners are also expected to 
contribute to $B \bar{B}$ mixing, changing the pattern of the CP 
asymmetries in B meson decays. Grand unified theories with a flavor U(2) 
symmetry give more complete and predictive theories of fermion masses.

An important consequence of the U(2) symmetry is that the symmetry 
structure of neutrino masses becomes similar, but not identical, to that 
of the charged fermion masses. The similarities ensure that predictions 
can be made, while the differences lead to an unusual result from the 
seesaw mechanism. 

The $U(2)$ theories of flavor are based on three assumptions:

\noindent {\em i}.) {\it Flavor physics is governed ultimately by a flavor 
symmetry $U(3)$, under which the three generations transform as a {\bf 3}.}

This assumption identifies the flavor space as the horizontal space of the
three generations. The flavor group acts identically on all charge 
components of a generation. This simple assumption follows directly from
theories having a unified vertical gauge symmetry, but can also occur in
non-unified theories. This assumption greatly constrains the flavor structure
of theories; for example, charged fermion masses transform as ${\bf \bar{3}} +
{\bf 6}$, while Majorana neutrino masses transform as ${\bf 6}$ -- 
there is a crucial connection between charged and neutral fermion masses.
In this paper we study theories containing right-handed neutrinos. They 
are assumed to be part of the generations so that they have Majorana 
masses which transform as ${\bf 6}$, and Dirac masses which transform as 
the charged fermion masses.

\noindent {\em ii}.) {\it $U(3)$ is broken strongly to $U(2)$ in all 
charged sectors.}

The large mass of the top quark is a signal that the $U(3)$ symmetry is
strongly broken, by couplings of order unity, to $U(2)$. We assume that this
large breaking is also manifest in the other charged sectors\footnote{In 
view of the lightness of the $b$ quark and $\tau$ lepton, relative to 
the $t$ quark, this assumption could be questioned. However, we take the 
view that there is an overall suppression of the down quark and charged 
lepton masses due to effects in the Higgs sector.}. The three generations 
transform as ${\bf 2} + {\bf 1}$: $\psi_a + \psi_3$. The entries in the 
Dirac fermion mass matrices therefore transform as: $\psi_3 \psi_3, 
\psi_3 \psi_a, \psi_a \psi_b$, and so are generated by the vevs of 
fields which transform as $\phi^a, S^{ab}$ and $A^{ab}$, where $S$ 
and $A$ are symmetric and antisymmetric tensors respectively.

Since $U(2)$ has rank 2, it can be broken in two stages. The only 
breaking pattern which leads to a hierarchy of masses for the three 
generations is
$$
U(2) \stackrel{\epsilon}{\rightarrow} 
U(1) \stackrel{\epsilon'}{\rightarrow} 0
$$
where $\epsilon$ and $\epsilon'$ are two small symmetry breaking parameters.

\noindent {\em iii.}) {\it  The vevs of all components of $\phi^a, S^{ab}$ 
and $A^{ab}$ are restricted to be of order $\epsilon$, $\epsilon'$ or 0.} 

This ensures that the magnitude of every entry in both the charged fermion
and neutrino mass matrices has a magnitude which is controlled by the $U(2)$ 
symmetry and its breaking.

The assumptions stated above, together with the phenomenological 
considerations discussed in Section~\ref{sec:three}, lead to definite
predictions in the U(2) model for neutrino mass ratios and mixing angles.
The neutrino physics of other viable supersymmetric flavor models can 
be found in the recent literature, for both Abelian \cite{abelian} and 
non-Abelian \cite{nonabelian} flavor groups.

\section{Canonical Models} \label{sec:two} \setcounter{equation}{0}

If we assume that a flavor symmetry $G_F$ acts identically on all 
members of a 16-plet generation, and that the symmetry is broken sequentially 
by a set of flavon fields $\{ \Phi_i\}$ that are symmetric under interchange 
of the matter fields, then we will obtain mass matrices for the quarks, 
charged leptons, and neutrinos that have identical textures, up to factors 
of order unity.  It will be instructive to consider the implications
for neutrino physics in this simple class of models before we move on 
to more complicated possibilities later.  In theories of this type, the 
differences between the up, down and charged lepton mass matrices must 
originate from fluctuations in the order one coefficients.  Such theories, 
however, are far from satisfactory.  While the down quark Yukawa couplings 
fall in the 
approximate ratio
\begin{equation}
h_d :: h_s :: h_b \approx \lambda^4 :: \lambda^2 :: 1  \,\,\, ,
\label{eq:rdown}
\end{equation}
the up quark Yukawa couplings are even more hierarchical
\begin{equation}
h_u :: h_c :: h_t \approx \lambda^8 :: \lambda^4 :: 1  \,\,\, ,
\label{eq:rup}
\end{equation}
where $\lambda \approx 0.22$ is the Cabibbo angle.
In order to explain the difference between Eqs.~(\ref{eq:rdown}) 
and (\ref{eq:rup}), some `order one' coefficients must 
differ from unity by more than one power of $\lambda$.
Hence, the fermion masses are not completely determined by the 
flavor symmetry breaking pattern and naive dimensional analysis.   In 
any realistic model where $G_F$ acts identically on the matter fields 
in a full generation, and where the `order one' coefficients 
really are near unity, flavor symmetry breaking in the up sector must 
occur at higher order.  As we will see later, this can happen if the 
flavons transform nontrivially under the grand unified group.

With this in mind, there are now a number of possibilities for the flavor
symmetry breaking pattern in the neutrino sector.  First, it is possible 
that each entry of the Dirac and Majorana neutrino mass matrices,
$M_{LR}$ and $M_{RR}$, will experience $G_F$ breaking at the same order as 
the corresponding entry of the down quark mass matrix $M_D$. 
In this case, all three mass matrices will have the same texture, while
the up quark mass matrix will differ due to some additional mechanism.
In a model of this type, the eigenvalues of  $M_{LR} \approx M_{RR}$ 
will be in the approximate ratio $\lambda^4 :: \lambda^2 :: 1$ and all 
mixing angles will be CKM-like.  For example, we might have
\begin{equation}
M_{LR}\sim M_{RR} \sim M_D \sim
\left(\begin{array}{ccc} \lambda^4 & \lambda^3 & \lambda^3 \\
                         \lambda^3 & \lambda^2 & \lambda^2 \\
                         \lambda^3 & \lambda^2 &   1  \end{array}
                         \right) \,\,\, ,
\label{eq:dlike}
\end{equation}
where $\sim$ indicates that we are interested only in the relevant 
hierarchies, and not in overall mass scales or in coefficients
of order unity. The left-handed Majorana mass matrix $M_{LL}$ is 
then given by the seesaw mechanism \cite{seesaw}
\begin{equation}
M_{LL} \approx M_{LR} \, M_{RR}^{-1} \, M_{LR}^T \,\,\, ,
\end{equation}
and we find
\begin{equation}
M_{LL}\sim 
\left(\begin{array}{ccc} \lambda^4 & \lambda^3 & \lambda^3 \\
                         \lambda^3 & \lambda^2 & \lambda^2 \\
                         \lambda^3 & \lambda^2 &   1  \end{array}
                         \right) \,\,\, .
\label{eq:intuitive}
\end{equation}
In this example, the left-handed neutrino spectrum has mass ratios and 
mixing angles that are comparable to those of the down quarks or charged 
leptons.   We will refer to models of this type as ``canonical".  In such
models, the neutrino mass matrices $M_{RR}$, $M_{LR}$, and $M_{LL}$ each 
have eigenvalues in the ratio $\lambda^4$ :: $\lambda^2$ :: $1$ and 
mixing angles bounded by the corresponding CKM angles, $\theta_{12}
\mbox{\raisebox{-1.0ex} {$\stackrel{\textstyle ~<~}
{\textstyle \sim}$}} \lambda$, $\theta_{13} 
\mbox{\raisebox{-1.0ex} {$\stackrel{\textstyle ~<~}
{\textstyle \sim}$}} \lambda^3$ and $\theta_{23} 
\mbox{\raisebox{-1.0ex} {$\stackrel{\textstyle ~<~}
{\textstyle \sim}$}} \lambda^2$.  Since canonical models have no order 
one mixing angles, they cannot explain the atmospheric neutrino deficit, nor
can they account for the large angle MSW or vacuum oscillation solutions 
to the solar neutrino problem \footnote{See Ref.~\cite{yan}, and
references therein.}.   The small angle MSW solution is possible 
in a canonical model only if the Cabibbo angle originates in the up 
quark sector, and the 12 mixing in $M_D$ is of order $\lambda^2$. 
  
More interesting results are obtained when $M_{LR}$, $M_{RR}$ and
$M_D$ have differing textures.   In theories where a full generation 
of the matter fields transforms identically under $G_F$, this may 
happen for two reasons:

{\em i}.) There are flavons in the theory that are purely antisymmetric 
under interchange of the matter fields.  These may contribute to
all the mass matrices except $M_{RR}$, which is purely symmetric.

{\em ii}.)  Other symmetries restrict the form of the mass matrices.
For example, some of the flavons may transform nontrivially under the 
grand unified group, $G_{GUT}$, so that the corresponding mass
matrix elements  are generated only after both $G_F$ and $G_{GUT}$ 
are broken.  This may produce the desired suppression of the 
up and charm quark masses, but may also lead to a suppression
of entries in the neutrino mass matrices as well.

In our previous example, it is simple to show that even a modest 
variation in the form of $M_{RR}$ away from Eq.~(\ref{eq:dlike}) can 
lead to bizarre results.  For example, we could imagine a theory 
where $M_{RR}$ has CKM-like mixing angles, but eigenvalues in the
ratio $\lambda^6$ :: $\lambda^4$ :: $1$, as follows from
\begin{equation}
M_{RR} \sim \left(\begin{array}{ccc}
\lambda^6 & \lambda^5 & \lambda^3 \\
\lambda^5 & \lambda^4 & \lambda^2 \\
\lambda^3 & \lambda^2 &     1  \end{array}\right)
\,\,\, .
\end{equation}
This form is obtained when the light two-by-two block is suppressed by
$\lambda^2$ compared to the canonical example of Eq.~(\ref{eq:dlike}).  
The seesaw mechanism now gives
\begin{equation}
M_{LL} \sim \left(\begin{array}{ccc}
\lambda^2 & \lambda & \lambda \\
\lambda &     1   &     1   \\
\lambda &     1   &     1    \end{array}\right) \,\,\, .
\label{eq:notint}
\end{equation}
Notice that the neutrino mass eigenvalues are in the ratio 
$\lambda^2$ :: $1$ :: $1$, and the mixing angles are not all CKM-like. 
This is a result that we would not have anticipated based on our knowledge of
flavor symmetry breaking in the down quark or charged lepton sector, and our
intuition alone.

In the remainder of this paper, we will consider the phenomenology
of models with a U(2) flavor symmetry \cite{u212,u23}.  In these models, 
complete generations transform identically under the flavor group, and 
either {\em i}, {\em ii}, or both are true, so that the neutrino
mass ratios and mixing angles are often noncanonical. In particular, we 
will see that a wide class of U(2) models predict order one
$\nu_\mu$-$\nu_\tau$ mixing, even though they involve no special
assumptions that would allow us to anticipate such a result.

\section{The Standard U(2) Model and a Simple Modification} \label{sec:three}
\setcounter{equation}{0}

Models with flavons in antisymmetric representations of the flavor
group may yield noncanonical neutrino mass ratios and mixing angles, even 
when this is the only factor that distinguishes the Dirac and Majorana 
neutrino mass matrices.  This fact is particularly significant in 
models with a U(2) flavor symmetry \cite{u212,u23}.  
In these models, the full 16-plet of matter fields $\psi$ transforms 
as a ${\bf 2}+{\bf 1}$ under U(2). Flavor symmetry breaking is 
achieved via three flavons,
\begin{equation}
S^{ab}, \,\,\, A^{ab}, \,\,\, \mbox{and} \,\,\, \phi^a ,
\end{equation}
where $S$ and $A$ are symmetric and antisymmetric tensors, respectively,
and $\phi$ is a doublet.  To obtain a viable texture for the down
quark Yukawa matrix, we require U(2) to be sequentially broken:
\begin{equation}
U(2) \longrightarrow U(1) \longrightarrow \mbox{nothing} \,\,\, ,
\label{eq:seqb}
\end{equation}
where the first stage of symmetry breaking is achieved via the vevs
\begin{equation}
\langle \phi^2 \rangle / M \approx \langle S^{22} \rangle / M 
= \epsilon \,\,\, ,
\label{eq:stageone}
\end{equation}
where $M$ is the flavor scale.  In the ``standard" U(2) model, the 
remaining U(1) symmetry is broken at a lower scale via the antisymmetric 
tensor, so that
\begin{equation}
\langle A^{12} \rangle / M = \epsilon' \,\,\, .
\label{eq:stagetwo}
\end{equation}
With this pattern of symmetry breaking, the down quark Yukawa matrix
has the texture
\begin{equation}
h_D \sim \left(\begin{array}{ccc}
0 & \epsilon' & 0 \\
-\epsilon' & \epsilon & \epsilon \\
0 & \epsilon & 1 \end{array}\right) \,\,\, ,
\label{eq:usual}
\end{equation} 
where we have omitted the order one coefficients. Eq.~(\ref{eq:usual})
yields acceptable mass ratios and mixing angles with $\epsilon \sim \lambda^2$
and $\epsilon'$ between $\lambda^3$ and $\lambda^4$.  Precise values for 
these parameters and the order one coefficients, obtained from a global fit, 
can be found in Ref.~\cite{u23}.  

The crucial issue that must be addressed in any realistic U(2) model is 
the origin of the differing mass hierarchies in the down and up quark
sectors, Eqs.~(\ref{eq:rdown}) and (\ref{eq:rup}).  The difference between the
top and bottom quark masses may be explained by a large value for the ratio of
Higgs vevs (i.e. $\tan \beta = \langle H_u \rangle / \langle H_d \rangle 
\sim 40$) or by an overall small parameter in $h_D$ originating from mixing in
the Higgs sector of the theory.   With the choice of $\epsilon$ 
and $\epsilon'$ given above, however, all the Yukawa matrices will have 
eigenvalue ratios that are characteristic of the down quarks.  Clearly, the 
sequential breaking of the flavor symmetry cannot account for 
the differing up and down quark mass hierarchies alone. Therefore, the 
transformation properties of the flavons under the grand unified 
group $G_{GUT}$, and perhaps also the orientation of the flavon vevs 
in GUT space must explain why the up and charm masses are generated 
at higher order in the flavor symmetry breaking. 

The precise mechanism that is responsible for suppressing $m_u$ and $m_c$ 
in U(2) theories is in fact a model-dependent question.  The relevant issue 
is whether this mechanism also affects the entries of the neutrino 
mass matrices, so that their sizes are not what we would expect naively 
from a sequential breaking of the U(2) symmetry.  In the next section, 
we will address this issue explicitly in the context of a well-motivated
effective theory, the SU(5)$\times$U(2) model.  We will find that a
suppression of some entries of the neutrino mass matrices does occur
and has interesting consequences.  In the remainder of this section,
however, we will consider the class of model in which the neutrino mass 
matrix elements are determined only by the scales of sequential U(2) 
breaking.  We will first comment briefly on U(2) models without
right-handed neutrinos, and then focus on the models of 
interest, which have complete 16-plet generations.

In U(2) models without right-handed neutrinos, the left-handed
Majorana mass matrix originates from a higher-dimension operator of 
the form $LHLH/ \Lambda$, where $\Lambda$ is some high scale,
perhaps the ratio of the Planck scale squared to the scale
where lepton number is violated.  The form of $M_{LL}$ is 
determined by the pattern of symmetry breaking in 
Eqs.~(\ref{eq:seqb}), (\ref{eq:stageone}), and (\ref{eq:stagetwo}), 
and we find
\begin{equation}
M_{LL} \approx \frac{H^2}{\Lambda} \left(\begin{array}{ccc}
0 & 0 & 0 \\
0 & \epsilon & \epsilon \\
0 & \epsilon &    1  \end{array} \right) \,\,\, .
\end{equation}
Notice that the 12 and 21 entries have vanished due to the antisymmetry 
of $A^{ab}$.  The muon and tau neutrinos have masses in the 
ratio $\epsilon$ :: $1$, while the electron neutrino is massless.  The 23 
mixing angle is of order $\epsilon \approx 0.02$, while the 13 mixing is 
negligible.  The 12 mixing angle originates from diagonalization of the 
charged lepton mass matrix, and is given by $\theta_{12} =  
\sqrt{m_e/m_\mu} \approx \epsilon'/(3 \epsilon) \approx 0.07$.  
This angle is too large by about a factor of 2 to yield the small
angle MSW solution to the solar neutrino problem, but may explain the 
LSND neutrino oscillation signal \cite{LSND} if the muon neutrino mass 
squared is in the range $0.3$-$0.6$ eV$^2$.

In considering U(2) models with right-handed neutrinos, we will also
assume here that both the Dirac and Majorana neutrino mass matrices have 
entries determined by the pattern of U(2) breaking, without any additional 
suppression.  An example of a theory of this type is the second
SO(10)$\times$U(2) model of Ref.~\cite{u23}, with all flavons transforming as
adjoints of SO(10), and a flavor-singlet ${\bf 126}$ added to generate the
right-handed neutrino scale.  In this model, the orientation of the flavon
vevs in GUT space assures a suppression of the lowest order contributions to
$m_u$ and $m_c$, but does not alter the form of the remaining Yukawa matrices,
when all operators are taken into account.  

In models of this type, the neutrino Dirac mass matrix $M_{LR}$ has 
the same form as $h_D$, while $M_{RR}$ is  given by
\begin{equation}
M_{RR} \approx \Lambda_R \left(\begin{array}{ccc}
0      &     0        &    0       \\
0      &  \epsilon    &   \epsilon \\
0      &  \epsilon    &      1     \end{array}\right)  \,\,\, ,
\label{eq:zeroe}
\end{equation}
where $\Lambda_R$ is the right-handed neutrino mass scale. 
The absence of a contribution from the antisymmetric flavon not
only has given us a different texture from $h_D$, but also has created a 
serious problem:  Eq.~(\ref{eq:zeroe}) has a zero eigenvalue.  If the 
seesaw mechanism is to be effective for all three generations, we must 
decide how to modify the theory (or our assumptions) so that all the 
eigenvalues of $M_{RR}$ are nonvanishing.  

A simple solution that does not require us to modify the field content of the 
theory, is to relax the assumption made in Refs.~\cite{u212,u23} that each
flavon is involved only in a single stage of the symmetry breakdown.  Thus, we
will consider the possibility that $S^{11}$, $S^{12}$, and $\phi^{1}$ have
nonvanishing vacuum expectation values of order $\epsilon'$.  Notice that 
these tensor components cannot acquire a vacuum expectation value until the
U(1) symmetry is spontaneously broken, so in general
\begin{equation}
S^{11} \mbox{\raisebox{-1.0ex} {$\stackrel{\textstyle ~<~}
{\textstyle \sim}$}} \epsilon', \,\,\,\,\,  S^{12} 
\mbox{\raisebox{-1.0ex} {$\stackrel{\textstyle ~<~}
{\textstyle \sim}$}} \epsilon',
\,\,\,\,\, \mbox{and} \,\,\,\,\, \phi^1 
\mbox{\raisebox{-1.0ex} {$\stackrel{\textstyle ~<~}
{\textstyle \sim}$}} \epsilon' \,\,\, .
\end{equation}
We will assume that these relations are equalities, so that the 
size of every nonvanishing Yukawa matrix element is set by one of the 
possible scales of sequential U(2) breaking.  The alternative, that $S^{11}$, 
$S^{12}$ and $\phi^{1}$ develop vevs at scales far below the U(1) breaking 
scale, yields Yukawa textures that cannot be understood solely in terms of 
a symmetry breaking pattern.   Since this possibility leads to less 
predictivity in the neutrino sector, we will consider it separately 
in the Appendix.

If we allow $S^{11}$, $S^{12}$ and $\phi^1$ to be either $0$ or
${\cal O}(\epsilon')$, we must consider the effects of our choice on the
phenomenology of the quark and charged lepton sectors:
 
$\bullet$ $S^{11}\approx \epsilon'$.   This leads to a texture for
the down-strange Yukawa matrix
\begin{equation}
\left(\begin{array}{ccc}
\epsilon' & \epsilon' \\
-\epsilon' & \epsilon \end{array}\right) \,\,\, ,
\end{equation}
which implies that the Cabibbo angle $\theta_c$ is given approximately
by $m_d/m_s$.  The measured value of the $\theta_c$ is described
quite accurately by $\sqrt{m_d/m_s}$, so our result is not
phenomenologically acceptable.  Thus we must choose $S^{11} = 0$. 

$\bullet$ $S^{12}\approx \epsilon'$.  Since the only antisymmetric
flavon in the theory contributes to the 12 entries of the Yukawa matrices,
the choice $S^{12}\approx \epsilon'$ guarantees that no Yukawa entry
is dominated by the contribution of an antisymmetric flavon.  Thus,
we obtain a canonical model, with mass ratios and mixing angles similar 
to those in the charged lepton sector.  With both symmetric and 
antisymmetric flavons present in the theory, the 12 and 21 entries 
of $h_U$ and $h_D$ no longer have a definite symmetry under interchange, and 
the successful prediction of the original theory $\theta_c = 
\sqrt{m_d/m_s}$ is violated at the 100\% level.  Since this relation is known
to be valid within 20\%, taking into account the 
allowed range $m_s/m_d = 17\mbox{ to }25$ \cite{pdg}, we conclude that 
theories with both $A^{12}$ and $S^{12}$ nonvanishing are not favored.
Thus, we are led to consider $S^{11}=S^{12}=0$ and $\phi^1 \approx \epsilon'$
as the most promising U(2) breaking pattern for both the neutral and 
charged fermion masses.

$\bullet$ $\phi^1\approx \epsilon'$.  Notice that adding 13 and 31 
entries to $h_D$ of order $\epsilon'$ corrects the down quark mass at 
the percent level, which is negligible.  However, there are now new 
contributions to $m_{u}$ and $V_{ub}$ that are of the same order as the 
ones in the original theory.  The only predicted relation involving these 
observables that is known accurately enough to be affected significantly by 
these new contributions is $V_{ub}/V_{cb} = \sqrt{m_u / m_c}$.  Since  
$V_{ub}/V_{cb} = 0.08 \pm 0.02$, a 50\% correction to this
relation would be within the range allowed at the 95\% confidence
level. Thus, if the $\phi^1$ vev is slightly smaller than $\epsilon'$, 
say $\epsilon'/3$, then the only effect on the phenomenology 
of the quark sector would be to alter some of the detailed predictions 
of the $\phi^1=0$ theory, obtained via a global fit in Ref.~\cite{BHR97}.
The phenomenological viability of the model, however, would not be 
affected.

In light of these arguments, we will adopt the choice $\phi^1 \sim \epsilon'$,
$S^{11}=S^{12}=0$, and proceed with the analysis of the neutrino sector.  
A somewhat smaller choice for $\phi^1$ will not affect the form of our 
results, which are only valid up to order 1 factors.  The right-handed 
neutrino mass matrix is now given by
\begin{equation}
M_{RR} = \Lambda_R \left(\begin{array}{ccc}
0         & 0 & \epsilon' \\
0         & \epsilon & \epsilon \\
\epsilon' & \epsilon &    1 \end{array} \right) \,\,\, ,
\end{equation}
and the seesaw mechanism gives 
\begin{equation}
M_{LL} = \frac{H^2}{\Lambda_R} \left(\begin{array}{ccc}
\epsilon'^2/\epsilon  &  \epsilon' & \epsilon' \\
\epsilon' &   1 & 1 \\
\epsilon' &   1 & 1 \end{array}\right)  \,\,\, .
\label{eq:firstres}
\end{equation}
Note that we have not included operators involving the flavon
product $\phi^a\phi^b$ for simplicity.  If these operators are present,
it is straightforward to check that they have no effect on the form of our 
result\footnote{Note that the results presented here and in the next
section remain unchanged in form by the field redefinitions 
required to place the neutrino kinetic terms in canonical form after 
the small U(2)-breaking corrections to the Kahler potential are taken into
account.}. The interesting feature of
Eq.~(\ref{eq:firstres}) is the order 1 mixing in the 2-3 block, which allows
for a possible solution to the atmospheric neutrino problem, via
$\nu_\mu$-$\nu_\tau$ oscillation.  The preferred parameter range for this
solution, $\delta m^2_{23} \approx 10^{-2 \pm 0.5}$ eV$^2$ 
and $\sin^2 2\theta_{23} \approx 0.4$ -- $0.6$ \cite{yan}, may be obtained by 
appropriate choices for $\Lambda_R$ and the order $1$ 
coefficients\footnote{One might worry that a $\phi^1$ vev somewhat
smaller than $\epsilon'$ might alter our conclusion that the $2$-$3$
mixing angle is of order one. Let us assume that the $\phi^1$ vev 
is $a \epsilon'$ and that the operator involving the antisymmetric flavon
that contributes to $M_{LR}$ has a coefficient $b$.  Then the 2-3 block  
of Eq.~(\ref{eq:firstres}) scales as
\begin{equation}
\left(\begin{array}{cc} (b/a)^2 & (b/a) \\
                        (b/a)   &  1    \end{array}\right)
\end{equation}
Any systematic deviation away from the order $1$ entries in
Eq.~(\ref{eq:firstres}) due to a slightly smaller choice for
the $\phi^1$ vev can be compensated by a slightly smaller
choice for the coefficient $b$.  Thus, we obtain the large
$2$-$3$ mixing angle without a significant fine-tuning.}.  
Neutrino oscillations in this parameter range would be observable 
at proposed long-baseline experiments, such as the 
KEK-SuperKamiokande, MINOS, or CERN-ICARUS experiments \cite{parke}.
The 13 mixing angle in Eq.~(\ref{eq:firstres}) is of order $\epsilon' \approx
\lambda^3$--$\lambda^4$, and is unlikely to have measurable consequences if
the overall neutrino mass scale is determined by the atmospheric neutrino
deficit.  

The 12 mixing angle, however, is actually larger than $\epsilon'$ 
since it originates at leading order from the diagonalization of the 
charged lepton mass matrix.  Thus, we know $\theta_{12}$ quite 
accurately,
\begin{equation}
\theta_{12} = \sqrt{m_e/m_\mu} \,\,\, ,
\end{equation}
or $\sin^2 2\theta_{12} \approx 0.02$.  This may be large enough
to allow $\nu_\mu$-$\nu_\tau$ and $\nu_\mu$-$\nu_e$ oscillations
to be observed simultaneously at least at some of the long-baseline 
experiments mentioned above.  The neutrino mixing matrix $U$, defined 
by $\nu_{mass} = U \nu_{flavor}$, is given approximately by the product 
of successive two dimensional rotations in the 23 and 12 subspaces.  
Thus, neglecting CP violation, we obtain the simple form 
\begin{equation}
U = \left(\begin{array}{ccc}1 & -s_{12} & 0 \\
s_{12} c_{23} & c_{23} & s_{23}  \\
-s_{12} s_{23} & -s_{23} & c_{23}
\end{array} \right) \,\,\, ,
\end{equation}
where $c_{ij}$ ($s_{ij}$) is the cosine (sine) of the $ij$ mixing
angle.  The $\nu_\mu$-$\nu_e$ oscillation probability is then
given by
\begin{equation}
P(\nu_\mu \rightarrow \nu_e) = \sin^2 2\theta_{12} \left(
c_{23}^2 \sin^2 \delta_{12} t + s_{23}^2 \sin^2 \delta_{13} t
- s_{23}^2 c_{23}^2 \sin^2 \delta_{23} t \right)  \,\,\, ,
\label{eq:pmue1}
\end{equation}
where $\delta_{ij}=(m_i^2-m_j^2)/4E$, and $E$ is the beam energy.  
If we set $\theta_{23}$ to the central value suggested by the 
atmospheric neutrino anomaly, then Eq.~(\ref{eq:pmue1}) may be
written
\begin{equation}
P(\nu_\mu \rightarrow \nu_e) = 
0.0171 \sin^2 \delta_{12} t + 0.0029 \sin^2 \delta_{13} t
- 0.0025 \sin^2 \delta_{23} t  \,\,\, .
\label{eq:pmue2}
\end{equation}
The MINOS experiment is expected to measure the $\nu_\mu$-$\nu_e$ 
oscillation probability to an accuracy of $0.0044$ \cite{minos}, and
the ICARUS experiment may achieve a comparable sensitivity \cite{parke}.  
Thus, we have hope of measuring the first term in Eq.~(\ref{eq:pmue2}), 
which might provide a 2 sigma signal if the $\sin^2 \delta_{12} t$ factor is 
approximately $1/2$.  While this factor depends effectively on one 
free parameter, the muon neutrino mass, the amplitude of this term,
$A = \sin^2 2\theta_{12} \cos^2\theta_{23}$, is a fixed prediction of 
the theory.  For $\sin^2 2\theta_{23}$ in the range $0.4$-$0.6$, 
$A$ must fall in the range
\begin{equation}
0.016 \leq  A  \leq 0.018  
\end{equation}
if the models presented in this section are correct.  Since the 
muon neutrino oscillates primarily to $\nu_\tau$ in our model, one 
might worry that this small $\nu_e$ oscillation signal would be swamped
by the background electrons coming from $\tau$ decays.  
Fortunately, these electrons have a softer energy spectrum than those 
produced directly via $\nu_e$ charged current scattering.  Thus, the 
$\nu_\mu$-$\nu_e$ oscillation signal may be isolated by placing an 
appropriate cut on the electron energy spectrum \cite{minos}. 

Finally, on a more speculative note, it is possible that our model 
can also account for the $\overline{\nu}_\mu$-$\overline{\nu}_e$
oscillation signal reported by the LSND experiment \cite{LSND}.
Given our prediction that $\sin^2\theta_{12} \approx 0.02$, the LSND 
results favor a $\Delta m^2_{12}$ in the range $0.3$-$0.6$ eV$^2$ \cite{LSND}.
At face value, this mass scale seems too large to account for the 
atmospheric neutrino deficit, in the absence of a 10\% fine-tuning.  
However, the only obstacle to solving the atmospheric neutrino problem via 
$\nu_\mu$-$\nu_\tau$ oscillation with $\Delta m^2_{23}$ in a similar range 
is a bound coming from the observed flux of upward-going muons at the 
IMB experiment \cite{imb}.  The observed flux is roughly comparable to 
theoretical expectations, and can be used to exclude a region of
the $\sin^2 2\theta$-$\Delta m^2$ plane that overlaps with the
region preferred by the atmospheric neutrino data for $\Delta m^2_{23}$ 
larger than 0.03.  A possible loophole is that this bound depends sensitively 
on the absolute neutrino flux, which has a large theoretical uncertainty.  
The most optimistic estimates for this flux (from our point of view) yield 
no constraint on the region of parameter space favored by the atmospheric 
neutrino deficit, beyond those already available from other 
experiments \cite{imb}, and allow a solution with $\Delta m^2_{23}$ 
as large as $0.4$ eV$^2$ \cite{barish}.  This would be sufficient to 
explain both the atmospheric and LSND phenomena, without any fine-tuning.  
If this interpretation is correct, it would also imply that 
the $\nu_\mu$-$\nu_e$ mixing angle would lie only a factor of $2$ 
below the current bounds from reactor experiments~\cite{LSND}.

\section{The SU(5)$\times$U(2) Model} \label{sec:four}
\setcounter{equation}{0}

We have seen in the previous section that U(2) models with flavons in 
antisymmetric representations of the flavor group may yield textures for 
$M_{LL}$ that have noncanonical mass ratios and mixing angles.  Another 
factor that may contribute to deviations from the canonical result is 
an additional suppression of some of the flavor-symmetry-breaking operators 
due to the transformation properties of the flavon fields under 
the grand unified group.  This is a possibility we will take into account in 
this section.  We will work in the context of SU(5), which is contained
in all other grand unified groups.  In principle, the number 
of possible effective U(2) theories for neutrino masses grows considerably 
if we also allow the flavons to have nontrivial transformation
properties under SU(5).  However, we will argue (as in Ref.~\cite{u23}) that 
one particular set of quantum number assignments for the flavons seems favored
by the known phenomenological differences between the up, down, and charged 
lepton Yukawa matrices.  This will enable us to make specific predictions 
in the neutrino sector as well.

In the SU(5)$\times$U(2) model of Ref.~\cite{u23}, fermion masses originate
from the operators
\begin{equation}
T_3 H T_3 + T_3 \overline{H} \, \overline{F}_3
\label{eq:third}
\end{equation}
\begin{equation}
+\frac{1}{M} \left(T_3 \phi^a H T_a + T_3 \phi^a \overline{H} \,
\overline{F}_a + \overline{F}_3 \phi^a \overline{H} T_a \right)
\label{eq:twothree}
\end{equation}
\begin{equation}
+\frac{1}{M} \left(T_a(S^{ab}H + A^{ab}H)T_b + 
T_a(S^{ab}\overline{H} + A^{ab} \overline{H}) \overline{F}_b\right)
\,\,\, ,
\label{eq:onetwo}
\end{equation}
where $T$ and $\overline{F}$ are the ${\bf 10}$  and ${\bf \overline{5}}$ 
matter multiplets, while $H$ and $\overline{H}$ are the {\bf 5} and 
${\bf \overline{5}}$ Higgs fields.  If all the flavons were SU(5) 
singlets, then the Yukawa matrices $h^{U}$, $h^{D}$, and $h^{E}$ would 
have the same U(2) breaking texture, and we would have no explanation 
for the differing mass hierarchies in the down- and up-quark sectors.
Thus, the SU(5) transformation properties of the flavons must account
for the known differences between $h^{U}$, $h^{D}$, and $h^{E}$.

The Yukawa matrices for the first two generations of up and down quarks
originate from the first pair and last pair of terms in 
Equation~(\ref{eq:onetwo}), respectively.  The simplest way of obtaining 
the differing mass hierarchies in Eqs.~(\ref{eq:rdown}) and 
(\ref{eq:rup}) is to choose SU(5) 
transformation properties for $S^{ab}$ and $A^{ab}$ such that they 
contribute at leading order to $h^{D}$, but not to $h^{U}$.  The crucial 
observation is that
\[
10 \times \overline{5} = 5 + 45
\]
while
\[
10 \times 10 = \overline{5}_s + \overline{45}_a + \overline{50}_s 
\,\,\, .
\]                           
The representations that contain a Higgs doublet (the {\bf 5} and {\bf 45})
are distinguished in the up sector by their definite symmetry under
interchange of the two {\bf 10}'s.  Thus, if we choose SH$\sim${\bf 45}
and AH$\sim${\bf 5}, the up quark mass will vanish at leading order, while
a charm mass may originate via the nonvanishing (2,3) and (3,2) entries 
of $h^{U}$, as we describe below. To realize this scenario, the flavons $A$ 
and $S$ must transform as a {\bf 1} and ${\bf 75}$, respectively.  Any 
other choice for the transformation properties of $A$ and $S$ that 
allows $AH$ and $SH$ to contain the desired SU(5) representations, also 
yields undesired representations as well\footnote{For example, if $A$ where 
to transform as a ${\bf 24}$ then $AH$ would indeed contain a ${\bf 5}$,
but would also have a component transforming as a ${\bf 45}$.}.
Thus, the quantum number assignments for the symmetric and antisymmetric 
flavons are significantly restricted. In fact, there is additional evidence 
that the choice $A\sim {\bf 1}$ and $S\sim {\bf 75}$ is a compelling one. 
The products $S\overline{H}$ and $A\overline{H}$ then transform as 
a ${\bf \overline{45}}$ and ${\bf \overline{5}}$, respectively, leading to 
a factor of 3 enhancement in the (2,2) entry of $h^{E}$.  We then 
automatically obtain the Georgi-Jarlskog mass relations at the GUT scale:
\begin{equation}
m_e = \frac{1}{3} m_d \,\,\,\,\,\,\,\,\,\,\,\, m_\mu = 3 m_s  \, .
\end{equation}
Therefore, we will assume $S\sim {\bf 75}$ and $A \sim {\bf 1}$ in our
subsequent analysis.  The remaining doublet flavons $\phi^a$ are needed to
generate the mixing between the second and third generations, and therefore 
must contribute to either (or both) the up and down sectors at lowest
order in the flavor symmetry breaking.  Since the components of the
Yukawa matrices generated by $\phi$ have no definite symmetry under
interchange of the matter fields, we expect $\phi$ to contribute
to both the up and down quark sectors, regardless of their SU(5) 
transformation properties.  A viable model is obtained with the minimal 
choice $\phi \sim {\bf 1}$, which we will assume henceforth.
If there are additional doublets in the theory that transform 
nontrivially under SU(5), their effects will be no larger than the SU(5) 
singlet contribution, and will not alter our results.  Notice that $\phi$ 
contributes to the (2,3) and (3,2) entries of $h^{U}$ at lowest order, so 
we generate a charm mass $m_c \sim \epsilon^2\sim\lambda^4$, as desired.  

Given these quantum number assignments, all the masses and 
mixing angles of the standard model are obtained, with the exception 
that the up quark is massless, $m_{u} = 0$, as a consequence of the 
combined grand unified and flavor symmetries.   An up quark mass can be 
generated at higher order, however, if we introduce additional 
fields \cite{u23}.  Let us suppose
that we also have a flavor-singlet, SU(5) adjoint field, whose vev points
in the hypercharge direction, $\Sigma_Y$.  This is the smallest
representation whose vev can break SU(5) down to the standard model
gauge group.  Then at order $1/M^2$, we have the operators
\begin{equation}
\frac{1}{M^2}\left(T_a \phi^a \phi^b H T_b + T_a S^{ab} \Sigma_Y H T_b
+ T_a A^{ab} \Sigma_Y H T_b \right)\,\,\, ,
\label{eq:muops}
\end{equation}
which always generate an up quark mass via the second and third terms.
To obtain an up Yukawa coupling of the appropriate magnitude, we find 
that $\Sigma_Y/M \approx \epsilon$, which is exactly the vev that
we would have expected based on dimensional analysis:  Since $S^{22} 
\approx \epsilon$, and $S$ transforms nontrivially under SU(5), we know 
that the flavor scale is approximately $1/\epsilon$ times higher than 
the unification scale.  For any purely SU(5) breaking vev $v$, we estimate 
that $v/M \approx \epsilon$, which is exactly what we need to
generate $m_{u}$ via the operators in Eq.~(\ref{eq:muops}).  Note
in addition that the second operator gives us another contribution to
$m_{c}$ that is of order $\lambda^4$.

Since we have found that the possible variations on the basic
SU(5)$\times$U(2) effective theory are significantly restricted, we 
have some hope for predictivity in the neutrino sector
\footnote{The SU(5) theory without right-handed neutrinos has a 
phenomenology identical to the corresponding theory in 
Section~\ref{sec:three}, except that the the muon and tau neutrino 
masses fall in the ratio $\epsilon^2$ :: $1$.  Therefore, we do not
discuss this case in the text.}. The Majorana mass matrix for the 
right-handed neutrinos $M_{RR}$ is generated at leading order by the  
operators
\begin{equation}
\Lambda_R \left(\nu_3 \nu_3 + \frac{1}{M} \phi^a \nu_a \nu_3
+ \frac{1}{M^2} \phi^a \phi^b \nu_a \nu_b
+ \frac{1}{M^3} S^{ab} \Sigma_Y \Sigma_Y \nu_a \nu_b \right)
\,\,\, ,
\end{equation}
where the two factors of $\Sigma_Y$ in the fourth term are
necessary to form an SU(5) singlet.  As in the model presented in
Section~\ref{sec:three}, we assume that $\phi^1 \approx \epsilon'$  
so that we lift the zero eigenvalue in $M_{RR}$ without spoiling the most
successful phenomenological predictions in the quark sector.  We
then obtain 
\begin{equation}
M_{RR} = \Lambda_R \left( \begin{array}{ccc}
\epsilon'^2 & \epsilon\epsilon' & \epsilon' \\
\epsilon\epsilon' &\epsilon^2   & \epsilon \\
\epsilon'  &  \epsilon & 1   
\end{array} \right) \,\,\, .
\end{equation}
Similarly, the neutrino Dirac mass matrix is generated by the operators
\[
\overline{F}_3 H \nu_3 + \frac{1}{M} (\phi^a \overline{F}_3 H \nu_a
+\phi^a \overline{F}_a H \nu_3 + A^{ab} \overline{F}_a H \nu_b)
\]\begin{equation}
+ \frac{1}{M^2} (\phi^a \phi^b \overline{F}_a H \nu_b
+ S^{ab} \Sigma_Y  \overline{F}_a H \nu_b)
\end{equation}
leading to the texture
\begin{equation}
M_{LR} = H \left( \begin{array}{ccc}
\epsilon'^2 & \epsilon' & \epsilon' \\
- \epsilon' & \epsilon^2 & \epsilon \\
\epsilon'& \epsilon & 1 \end{array} \right) \,\,\, .
\end{equation}
The seesaw mechanism then give us solutions with two
possible textures:
\begin{equation}
M_{LL} = \frac{H^2}{\epsilon \Lambda_R} \left(
\begin{array}{ccc}
(\epsilon'/\epsilon)^2 & \epsilon'/\epsilon & \epsilon\epsilon' \\
\epsilon'/\epsilon &   1  &  \epsilon \\
\epsilon \epsilon' & \epsilon  &  \epsilon  \end{array} \right)
\label{eq:su5one}
\end{equation}
if there is a single doublet flavon in the theory, or
\begin{equation}
M_{LL} = \frac{H^2}{\epsilon \Lambda_R} \left(
\begin{array}{ccc}
(\epsilon'/\epsilon)^2 & \epsilon'/\epsilon & \epsilon'/\epsilon \\
\epsilon'/\epsilon &   1  &  1 \\
\epsilon'/\epsilon &   1  &  1  \end{array} \right)
\label{eq:su5two}
\end{equation}
if there are two or more doublet flavons.   The smaller entries in
Eq.~(\ref{eq:su5one}) result from a cancellation in leading terms
due to the proportionality between the entries generated by a single 
$\phi^a$ in $M_{RR}$ and $M_{LR}$.  In the second case, we obtain  
${\cal O}(1)$ mixing between the second and third generation 
neutrinos, as in the model of Section~\ref{sec:three}.  The significant 
difference in this case is that the additional suppression of the operators 
involving the symmetric flavon $S^{ab}$ has yielded an enhancement in 
the 12 and 13 mixing angles, which are now both of order 
$\epsilon'/\epsilon \approx \lambda$.   Unlike our earlier models,
which had negligible 13 mixing, the neutrino mixing matrix $U$ in the 
present case does not assume a particularly simple form.  Moreover, the $12$ 
mixing angle comes primarily from the diagonalization of $M_{LL}$, and 
therefore is known only up to an order one factor.   While these results 
prevent us from achieving the (rather surprising) level of predictivity 
that we obtained in the models of Section~\ref{sec:three}, it is 
significant consolation that the $\nu_\mu$-$\nu_e$ mixing probability is  
nearly an order of magnitude larger in the present model.  If the neutrino 
mass scale is the proper one to solve the atmospheric neutrino problem, then 
it seems very likely in this model that $\nu_\mu$-$\nu_\tau$ and 
$\nu_\mu$-$\nu_e$ oscillations would be observed together at long-baseline 
neutrino oscillation experiments, assuming the anticipated sensitivity 
of the MINOS experiment.

\section{Conclusions}
\setcounter{equation}{0}
   
We have considered the implications of a non-Abelian flavor symmetry
on neutrino masses and mixing angles.  In models where complete
generations transform identically under the flavor symmetry, we
argued that neutrino mass matrix textures can differ dramatically 
from those of the charged leptons.  This may happen if there are
flavons in the theory that transform nontrivially under $G_{GUT}$, or that 
are antisymmetric under $G_F$. In theories with a U(2) flavor symmetry, we 
found that some noncanonical models predict a large $2$-$3$ mixing angle, 
and therefore may provide a natural solution to the atmospheric neutrino 
problem.  Assuming that this consideration sets the mass scale for 
the muon neutrino, the $\nu_\mu$-$\nu_e$ mixing angle in these models 
is large enough to be measured at proposed long-baseline neutrino 
oscillation experiments.  We would then expect $\nu_\mu$-$\nu_\tau$ and 
$\nu_\mu$-$\nu_e$ oscillations to be observed simultaneously, with 
events falling in the approximate ratio $1$ :: $0.02$ in the 
models of Section~\ref{sec:three}, or $1$ :: $0.1$ in the 
model of Section~\ref{sec:four}.

\begin{center}               
{\bf Acknowledgments} 
\end{center}
This work was supported in part by the Director, Office of 
Energy Research, Office of High Energy and Nuclear Physics, Division of 
High Energy Physics of the U.S. Department of Energy under Contract 
DE-AC03-76SF00098.  LJH was also supported in part by the National 
Science Foundation under grant PHY-95-14797.

\appendix
\section{Other Models}
\setcounter{equation}{0}

In the text, it was assumed that each nonvanishing entry of the
Yukawa matrices was associated with one of the scales of sequential U(2)
breaking.  The possible values for these matrix elements were
then given by $\epsilon$, $\epsilon'$, or $0$.  This assumption was 
particularly important in determining the flavon vevs needed to 
lift the zero eigenvalue in Eq.~(\ref{eq:zeroe}).  In this appendix, we 
point out that smaller vevs for $S^{11}$, $S^{12}$, and $\phi^1$ lead 
to neutrino mass spectra with a distinct qualitative feature -- a 
heavy, nearly decoupled muon neutrino mass eigenstate.  The generalization 
of our previous analysis is straightforward, and we work with the model 
of Section~\ref{sec:three} for the purposes of illustration.  If 
we assume that the $S^{11}$, $S^{12}$, and $\phi^1$ vevs are of order 
$\delta_1$, $\delta_2$, and $\delta_3$, respectively, then the Dirac 
and Majorana neutrino mass matrices become
\begin{equation}
M_{LR}=\left(\begin{array}{ccc}
\delta_1 &  \epsilon' + \delta_2 & \delta_3 \\
-\epsilon'+\delta_2  & \epsilon & \epsilon \\
\delta_3 & \epsilon & 1 \end{array} \right) \,\,\, ,
\label{eq:deltalr}
\end{equation}
\begin{equation}
M_{RR}=\left(\begin{array}{ccc}
\delta_1 &  \delta_2 & \delta_3 \\
\delta_2  & \epsilon & \epsilon \\
\delta_3 & \epsilon & 1 \end{array} \right) \,\,\, ,
\label{eq:deltarr}
\end{equation}
and again $M_{LL} = M_{LR} \, M_{RR}^{-1} \, M_{LR}^T$.  The general
form for $M_{LL}$ can be easily computed, but is somewhat cumbersome, 
so we will not display it explicitly.  However, the important qualitative 
result is easy to appreciate by considering some simplifying limits.

\noindent $\delta_1 \geq \delta_2,\delta_3$:
\begin{equation}
M_{LL}=\frac{H^2}{\Lambda_R}\left(\begin{array}{ccc}
\epsilon'^2/\epsilon &  \epsilon' & \epsilon' \\
\epsilon'  & \epsilon'^2/\delta_1 & \epsilon \\
\epsilon' & \epsilon & 1 \end{array} \right)
\end{equation}
$\delta_2 \gg \delta_1,\delta_3$:
\begin{equation}
M_{LL}=\frac{H^2}{\delta_2\Lambda_R}\left(\begin{array}{ccc}
0 & \epsilon'^2 & 0 \\
\epsilon'^2  & \epsilon\epsilon'^2/\delta_2 & \epsilon\epsilon' \\
0 & \epsilon\epsilon' & \delta_2 \end{array} \right)
\end{equation}
$\delta_2 \sim \delta_3 \gg \delta_1$:
\begin{equation}
M_{LL}=\frac{H^2}{\delta_2\Lambda_R}\left(\begin{array}{ccc}
\epsilon'^2 \delta_{2} & \epsilon'^2 & \epsilon'\delta_2 \\
\epsilon'^2  & \epsilon\epsilon'^2/\delta_2 & \epsilon\epsilon' \\
\epsilon'\delta_2 & \epsilon\epsilon' & \delta_2 \end{array} \right)
\end{equation}
$\delta_3 \gg \delta_1,\delta_2$:
\begin{equation}
M_{LL}=\frac{H^2}{\delta_3\Lambda_R}\left(\begin{array}{ccc}
\epsilon'^2\delta_3/\epsilon &  \epsilon'^2 & \epsilon'\delta_3 \\
\epsilon'^2  & \epsilon'^2/\delta_3 & \epsilon' \\
\epsilon'\delta_3 & \epsilon' & \delta_3 \end{array} \right)
\end{equation}
In each case it was assumed that the $\delta_i$ were the smallest scales
in the problem.  Notice also in the last case that we recover 
Eq.~(\ref{eq:firstres}) when $\delta_3 \approx \epsilon'$.  While the 
general form for $M_{LL}$ implied by Eqs.~(\ref{eq:deltalr}) and 
(\ref{eq:deltarr}) does not allow us to make very definite statements 
about the phenomenology of the neutrino sector, we do see from these 
limiting cases that a widely split neutrino spectrum, with a heavy muon 
neutrino, is another possibility in U(2) models with an antisymmetric 
flavon.

\end{document}